\newcommand{\vpp}{\vec{p}\,'}

\documentstyle[preprint,revtex]{aps}
\begin{document}
\def\overlay#1#2{\setbox0=\hbox{#1}\setbox1=\hbox to \wd0{\hss #2\hss}#1%
\hskip -2\wd0\copy1}
\begin{title}
Chiral Limit of Nucleon Lattice Electromagnetic
Form Factors
\end{title}
\author{Walter Wilcox}
\begin{instit}
Department of Physics, Baylor University, Waco, TX 76798
\end{instit}
\moreauthors{Terrence Draper and Keh-Fei Liu}
\begin{instit}
Department of Physics, University of Kentucky, Lexington, KY 40506
\end{instit}
\vspace{0.4cm}
\begin{abstract}
We calculate electric and magnetic
form factors of protons and neutrons in quenched Monte Carlo
lattice QCD on a $16^3\times 24$ lattice
at $\beta = 6.0$ using Wilson fermions.
We employ a method which characterizes one of the nucleon
fields as a fixed zero-momentum secondary source.
Extrapolating the overall data set
to the chiral limit, we find acceptable fits
for either dipole or monopole forms
and extract proton and neutron magnetic moments, the magnitude of which
are $10$ to $15\%$ low compared to experiment.
In the extrapolation of the dipole fit of
the form factors, we find that the dipole
to nucleon mass ratio is about $7\%$ low compared to experiment.
In addition, we obtain positive values of the neutron electric form factor,
which, however, are poorly represented by a popular phenomenological form
at intermediate to small $\kappa$ values.
A zero-momentum technique for extracting hadron magnetic moments is
briefly discussed and shown to yield unrealistically small magnetic
moment values.
\end{abstract}
\newpage
\section{INTRODUCTION}
The techniques of Monte Carlo lattice QCD continue to be developed and
applied to numerous quantities of phenomenological interest. In particular,
electromagnetic form factors are a useful probe of hadron
internal structure.
The method used in
Ref.~1 characterizes one of the meson
interpolating fields as a zero-momentum secondary source.
Here, we apply this technique to the nucleon.
This allows us, in the final analysis of the quark propagators,
to reconstruct a number of
operator probes, among which are the various lattice axial
and vector currents.
We report here on our results using the conserved lattice vector
current; see Ref.~2 for preliminary results using an axial probe.
Using this technique with separate proton and neutron electric and
magnetic sources, we
extract all four of the nucleon electromagnetic form factors
at any desired lattice momentum transfer.
(For an independent implementation of this technique
as applied to the proton, see Ref.~3, where, however,
only the proton electric form factor is studied.)
This investigation for the nucleon is complementary to those in
Refs.~4 and 5 which fix the vector current, rather than the
particle field, as the secondary source
and analyze the full set of spin $1/2$ baryon form factors,
but only for a single momentum transfer value per quark mass value.
See also Ref.~6 for an introduction to
the formalism and Ref.~7 for some preliminary results of the
present investigation.

The purpose of this study is to begin
a comprehensive examination of the mass and momentum
dependence of nucleon form factors in the quenched approximation.
In our treatment, we will extrapolate
the {\it parameters} of the functional forms rather than individual
form factor {\it values}. This is required if contact with
various phenomenological forms for nucleon form factors is to be made.

We begin with an introduction to the formalism of lattice sources and
correlation functions. We then present our results for the electric and
magnetic form factors and their extrapolation to the chiral limit.
We close with a summary and some comments about the directions
of future lattice calculations.

\section{FORMALISM}
\subsection{Fundamentals}
In order to set the notation, we first give a fairly complete
catalog of the definitions and symbols used in this paper.
Our conventions follow Sakurai\cite{sak}, although we adopt the non-standard
representation gamma matrices:
\begin{equation}
 \{\gamma_\mu,~\gamma_\nu\} = 2\delta_{\mu,\nu},~~ \gamma_\mu^\dagger
 = \gamma_\mu ,
\end{equation}
\begin{eqnarray}
 \vec{\gamma}= \left( \begin{array}{cc}
                      0   & ~~~\vec{\sigma} \\
           \vec{\sigma}  & ~~~0  \end{array}\right) ,~~~~
 \gamma_4 = \left( \begin{array}{cc}
                      I   &  ~~~0  \\
                      0   &  ~~-I  \end{array}\right) ,~~~~
 \gamma_5 = i \left( \begin{array}{cc}
                      0   &  ~~-I  \\
                      I   &   ~~~0  \end{array}\right) .
\end{eqnarray}
We use the four-vector notation $b_\mu =
(\vec{b},ib_{0})$ , where $b_{0}$ is purely real, to ease the transition to
Euclidean space. Our discrete Minkowski fermion action is given by
\begin{equation}
S_F^{M}[\bar{\psi},\psi] = -\sum_{I,J}\bar{\psi}_I M_{I J}
 [U^{\dagger}]\psi_J,\label{eq:act1}
\end{equation}
where $I = \{x,\alpha ,a\}$ and $J = \{y,\beta,b\}$, which
are in the order $\{$space-time, Dirac, color$\}$. Flavor sums are
understood where appropriate.
The matrix $M_{I J}[ U^{\dagger}]$ is defined by
\begin{eqnarray}
&M_{I J}[ U^{\dagger}]
\equiv \delta_{I,J} - \kappa \displaystyle{\sum_{\mu}\left[ \left( 1-
\gamma_{\mu} \right)_{\alpha \beta}\left(U_{\mu}(x)\right)^{a b}
\delta_{x,y-a_{\mu}}\right.} \nonumber\\
&\left. +\left( 1 + \gamma_{\mu} \right)_{\alpha
\beta}\left(U_{\mu}^{\dagger}(x-a_{\mu}) \right)^{a b}
\delta_{x,y+a_{\mu}}\right].\label{eq:mij}\end{eqnarray}
Using $\kappa = 1/2(ma + 4)$ and (all fields not otherwise specified
are lattice fields and \lq $a$\rq\, is the lattice spacing)
\begin{equation}
\psi \rightarrow \frac{1}{\sqrt{2\kappa}}a^{3/2}\psi^{cont.},\label{eq:cor1}
\end{equation}
we find that with $U_{\mu}^{\dagger}(x)=e^{iaA_{\mu}(x)}$, this corresponds to
the continuum action:
\begin{equation}
	S_{F}^{cont.} =  - \int dx_{1}dx_{2}dx_{3}dt
	\left(\overline{\psi}^{cont.}(x)\left(
	\gamma_{\mu} \left(\partial_{\mu}-iA_{\mu}^{cont.}
(x)\right) + m \right)\psi^{cont.}(x)
\right).\label{eq:sf}
\end{equation}
The assigned
$(1 \pm \gamma_{\mu})$ structure in
Eq.~(\ref{eq:mij}) means that the upper components of the Dirac
equation propagate in the forward time direction in the static
(small $\kappa$) limit.
The conserved vector current\cite{karsten} from
Eqs.~(\ref{eq:act1}) and (\ref{eq:mij}) is
identified from ($\Delta_{\mu}\omega (x)
\equiv \omega (x+a_{\mu})-\omega (x)$)
\begin{eqnarray}
&\psi (x) \rightarrow e^{-i\omega (x)}\psi (x) ,\\
&\overline\psi (x) \rightarrow \overline\psi (x) e^{i\omega (x)},\\
&J_{\mu} \equiv\displaystyle{
\frac{\delta S_F}{\delta\left[\Delta_{\mu}\omega (x)\right]}}.
\end{eqnarray}
$J_{\mu}= ( \vec{j},i\rho )$ is given explicitly by (for a single flavor)
\begin{equation}
	J_{\mu}(x)  = i \kappa
	\left(
 \overline{\psi}(x+a_{\mu})\left(1+\gamma_{\mu}\right)
	U_{\mu}^{\dagger}(x)\psi(x) -\overline{\psi}(x)\left(1-\gamma_{\mu}\right)
	U_{\mu}(x)\psi(x+a_{\mu})\right). \label{eq:ju}
\end{equation}
$\rho(x)$ is normalized by ($t > t' > 0$)
\begin{equation}
\sum_{\vec{x}'}
\langle {\rm vac}|\psi_{\alpha}^{a}(x)\rho (x') \overline\psi_{\beta}^{b}(0)
|{\rm vac}\rangle =\langle {\rm vac}
|\psi_{\alpha}^{a}(x) \overline\psi_{\beta}^{b}(0)
|{\rm vac}\rangle .
\end{equation}
The interpolating fields we use for the proton are
\begin{eqnarray}
 \chi_{\alpha} (x) &=& \epsilon^{abc}\psi_\alpha^{(u) a}(x)\psi_\beta^{(u) b}
 (x) (\tilde{C})_{\beta\gamma}\psi_\gamma^{(d) c}(x), \label{eq:int1}\\
 \overline{\chi}_{\alpha'}(x)&=& -\epsilon^{a'b'c'}\overline{\psi}_
{\gamma'}^{(d) c'}
(\tilde{C})_{\gamma' \beta'} \overline{\psi}_{\beta'}^{(u) b'}(x)
\overline{\psi}_{\alpha'}^{(u) a'}(x),\label{eq:int2}
\end{eqnarray}
where $u \leftrightarrow d $ for the neutron.
(We assume $m_{u}=m_{d}$ throughout.) The charge conjugation matrix
$C=\gamma_2$ satisfies $
	\tilde{C}\gamma_{\mu}\tilde{C}^{-1}
	 = \gamma_{\mu}^{*}
$ where $\tilde{C}=C\gamma_5$.
We now continue to
Euclidean space ($t \rightarrow -it$, where $t$ remains real)
and use the integration formula\cite{soper}
\begin{equation}
\langle {\rm vac}|T(\psi_{\alpha}(-it_{A})\overline\psi_{\beta}(-it_{B})\ldots
)
|{\rm vac}\rangle = Z^{-1}\int d U d\overline\zeta d\zeta \,e^{-S_G^{E}
-S_F^{E}[\overline\zeta,\zeta]} \zeta_{\alpha}(t_{A})\overline\zeta (t_{B})
\ldots, \end{equation}
where the $\zeta,\overline\zeta$ are independent, totally anti-commuting
Grassmann integration variables and
$S_{G}^{E}$, $S_{F}^{E}[\overline\zeta,\zeta]$
are the Euclidean gluon, fermion
actions\cite{explain}. For example, on our lattice ($\alpha$ and $\beta$ are
generic indices) %
\begin{eqnarray}
&\displaystyle{ \int}  d \overline\zeta d \zeta\,
\zeta_{\alpha}\overline\zeta_{\beta}\,
e^{-\overline\zeta M \zeta}
= {\rm det}(M) S_{\alpha \beta},\\
&\displaystyle{ \int} d \overline\zeta d \zeta \,
\zeta_{\alpha} \overline\zeta_{\beta}
\zeta_{\gamma}\overline\zeta_{\delta}\,
e^{-\overline\zeta M \zeta}
= {\rm det}(M) ( S_{\alpha \beta}S_{\gamma \delta} -
 S_{\alpha \delta}S_{\gamma \beta} ),
\end{eqnarray}
where $S \equiv  M^{-1}$.

\subsection{Source technique}
For the proton two-point function,
we define (assuming $t > 0$ in the second line)
\begin{eqnarray}
&G_{pp}(t;\vec{p},\Gamma)  \equiv
 \displaystyle{\sum_{\vec{x}}}
	e^{-i\vec{p}\cdot\vec{x}}
\Gamma_{\alpha'\alpha}
\langle{\rm vac}|T\left(\chi_{\alpha}(x)
	\overline{\chi}_{\alpha'}(0)\right)|{\rm vac}\rangle \nonumber\\
 &= \displaystyle{\sum_{\vec{x}}}
e^{-i\vec{p}\cdot\vec{x}}\Gamma_{\alpha'\alpha}
\epsilon^{abc}(-\epsilon
^{a'b'c'})(\tilde{C})_{\beta\gamma}(\tilde{C})_{\gamma'\beta'} \nonumber\\
 &  \langle {\rm vac}|\psi_\alpha^{(u) a}(x)\psi_\beta^{(u) b}(x)
\psi_\gamma^{(d) c}(x)
\overline{\psi}_{\gamma'}^{(d) c'}(0)\overline{\psi}_{\beta'}^{(u) b'}
(0)\overline{\psi}_{\alpha^\prime}^{(u) a'}(0)|{\rm vac}
\rangle.\label{eq:two}
\end{eqnarray}
We will specify later the particular $\Gamma$-matrices
which we use here and in following equations.
Defining in Dirac space the quantity
 $\underline{Q} \equiv \left(\tilde{C}Q
\tilde{C}^{-1}\right)^{T}$ for an arbitrary matrix
$Q$, we have ( setting ${\rm det}(M)=1$)
\begin{eqnarray}
	&G_{pp}(t;\vec{p},\Gamma) =
	\displaystyle{\sum_{\vec{x}}}
	e^{-i\vec{p}\cdot\vec{x}}
	\epsilon^{abc}\epsilon^{a^{\/\prime}b^{\/\prime}c^{\/\prime}}
	\left({\rm tr}\left[\Gamma
		           S^{(u)aa'}(x,0)
	        \underline{S}^{(d)bb'}(x,0)
		           S^{(u)cc'}(x,0)\right]\right.
								\nonumber \\
	&  \left.
	+     {\rm tr}\left[\Gamma   S^{(u)aa'}(x,0)\right]
	      {\rm tr}\left[        \underline{S}^{(d)bb'}(x,0)
		               S^{(u)cc'}(x,0)\right]
	\right),
\end{eqnarray}
where a configuration average is understood and the trace is over Dirac
indices only. The corresponding {\em connected} Wick contractions
of the proton three-point function are given as
($\vec{q}=\vec{p}-\vec{p}~'$)%
\begin{eqnarray}
	 &G_{pJ_{\mu}p}(t_{2},t_{1};\vec{p},\vpp,\Gamma)
	  \equiv
	-i\displaystyle{ \sum_{\vec{x}_{2},\vec{x}_{1}}}
	e^{-i \vec{p}      \cdot\vec{x}_{2}}
        e^{i\vec{q}\cdot\vec{x}_{1}}
	 \Gamma_{\alpha'\alpha} \langle{\rm vac}|
		T\left(\chi_{\alpha}(x_{2})
	J_{\mu}(x_{1})\overline{\chi}_{\alpha'}(0)\right)|{\rm vac}\rangle
\nonumber \\
	 & = \displaystyle{ \sum_{\vec{x}_{2}}}
	e^{-i\vec{p}\cdot\vec{x}_{2}}
	\epsilon^{abc}\epsilon^{a'b'c'}
	\left({\rm tr}\left[
	\Gamma  q_{u}\hat{S}^{(u)aa'}(x_{2},0;t_{1},\vec{q},\mu)
	             \underline{S}^{(d)bb'}(x_{2},0)
		     S^{(u)cc'}(x_{2},0) \right] \right. \nonumber \\
	 &+ {\rm tr}\left[
	\Gamma  S^{(u)aa'}(x_{2},0)
        q_{d}\underline{\hat{S}}^{(d)bb'}(x_{2},0;t_{1},\vec{q},\mu)
		     S^{(u)cc'}(x_{2},0) \right] \nonumber\\
&+  {\rm tr}\left[	\Gamma  S^{(u)aa'}(x_{2},0)
	             \underline{S}^{(d)bb'}(x_{2},0)
		q_{u}\hat{S}^{(u)cc'}(x_{2},0;t_{1},\vec{q},\mu) \right] \nonumber \\
  &+ {\rm tr}\left[	\Gamma
q_{u}\hat{S}^{(u)aa'}(x_{2},0;t_{1},\vec{q},\mu)\right]
	             {\rm tr}\left[\underline{S}^{(d)bb'}(x_{2},0)
		     S^{(u)cc'}(x_{2},0) \right]  \nonumber \\
	 &+ {\rm tr}\left[
	\Gamma  S^{(u)aa'}(x_{2},0)\right]
        {\rm tr}\left[q_{d}
\underline{\hat{S}}^{(d)bb'}(x_{2},0;t_{1},\vec{q},\mu)
		     S^{(u)cc'}(x_{2},0) \right] \nonumber\\
&+ \left.{\rm tr}\left[	\Gamma  S^{(u)aa'}(x_{2},0)\right]
	             {\rm tr}\left[\underline{S}^{(d)bb'}(x_{2},0)
		q_{u}\hat{S}^{(u)cc'}(x_{2},0;t_{1},\vec{q},\mu)
			\right] \right).\label{eq:pro}
\end{eqnarray}
This is similar to the two-point function, except that each
quark propagator, $S$, has been replaced, in turn,
by $q_{f}\hat{S}$ where $q_{f}$ is the quark charge
( $q_{u}=2/3, q_{d}=-1/3$) and
\begin{eqnarray}
	&\hat{S}(x_{2},0;t_{1},\vec{q},\mu) \equiv \kappa
	\displaystyle{ \sum_{\vec{x}_{1}}}
 e^{i\vec{q}\cdot\vec{x}_{1}}
 \left( S(x_{2},x_{1}+a_{\mu})(1+\gamma_{\mu})
	U_{\mu}^{\dagger}(x_{1})S(x_{1},0) \right. \nonumber \\
	& -\left.  S(x_{2},x_{1})(1-\gamma_{\mu})
	U_{\mu}(x_{1})S(x_{1}+a_{\mu},0) \right),\label{eq:shat}
\end{eqnarray}
describes the quark propagator coupled, with momentum $\vec{q}$,
to the electromagnetic current given in Eq.~(\ref{eq:ju}). In order to
compactify our notation, let us introduce the quantity
\begin{eqnarray}
\left(S_{A}(x_{2},0)\right)_{\alpha'\alpha}^{a'a} &\equiv
\epsilon^{abc}\epsilon^{a'b'c'}\left[\left(\underline{S}^{bb'}(x_{2},0)
S^{cc'}(x_{2},0)\Gamma\right)
_{\alpha'\alpha}+\left(\Gamma S^{bb'}(x_{2},0)
\underline{S}^{cc'}(x_{2},0)\right)
_{\alpha'\alpha} \right.\nonumber \\
&+ \left. {\rm tr}\left[\underline{S}^{bb'}(x_{2},0) S^{cc'}(x_{2},0)\right]
\Gamma_{\alpha'\alpha}+{\rm tr}\left[\Gamma S^{bb'}(x_{2},0)\right]
\left(\underline{S}^{cc'}\right)_{\alpha'\alpha}\right],
\end{eqnarray}
corresponding to the $u$ ($d$) quark contribution
to the three-point function for the proton (neutron), as well as
\begin{eqnarray}
\left(S_{B}(x_{2},0)\right)_{\alpha'\alpha}^{a'a} &\equiv
\epsilon^{abc}\epsilon^{a'b'c'}
\left[\left(\underline{S}^{bb'}(x_{2},0)\underline{\Gamma}\,
\underline{S}^{cc'}(x_{2},0)\right)_{\alpha'\alpha}\right. \nonumber \\
&+ \left. {\rm tr}\left[\Gamma S^{bb'}(x_{2},0)\right]
\left(\underline{S}^{cc'}(x_{2},0)
\right)_{\alpha'\alpha}\right],
\end{eqnarray}
corresponding to the $d$ ($u$) quark contribution to the proton
(neutron). In addition, define
\begin{equation}
\hat{S}(x_{2},0;t_{1},\vec{q},\mu) \equiv \sum_{y}S(x_{2},y)
X(y,0;t_{1},\vec{q},\mu),
\end{equation}
where $X(y,0;t_{1},\vec{q},\mu)$ is given explicitly as
\begin{eqnarray}
X(y,0;t_{1},\vec{q},\mu) = \kappa\displaystyle{ \sum_{\vec{x}_{1}}}
 e^{i\vec{q}\cdot\vec{x_{1}}}\left[
\delta_{y,x_{1}+a_{\mu}}
(1+\gamma_{\mu})U_{\mu}^{\dagger}
(x_{1}) S(x_{1},0) \right. \nonumber \\
- \left. \delta_{y,x_{1}}(1-\gamma_{\mu})
U_{\mu}(x_{1}) S(x_{1}+a_{\mu},0)\right].\label{eq:x}
\end{eqnarray}
Then we may write concisely for the proton and neutron
\begin{eqnarray}
G_{pJ_{\mu}p}(t_{2},t_{1};\vec{p},\vpp,\Gamma) =
q_{u}A_{J_{\mu}}(t_{2},t_{1};\vec{p},\vpp,\Gamma)
+ q_{d}B_{J_{\mu}}(t_{2},t_{1};\vec{p},\vpp,\Gamma), \label{eq:gpp}\\
G_{nJ_{\mu}n}(t_{2},t_{1};\vec{p},\vpp,\Gamma) =
q_{d}A_{J_{\mu}}(t_{2},t_{1};\vec{p},\vpp,\Gamma)
+ q_{u}B_{J_{\mu}}(t_{2},t_{1};\vec{p},\vpp,\Gamma), \label{eq:gnn}
\end{eqnarray}
where
\begin{eqnarray}
A_{J_{\mu}}(t_{2},t_{1};\vec{p},\vpp,\Gamma) \equiv
\sum_{\vec{x}_{2},y}e^{-i\vec{p}\cdot\vec{x_{2}}}
{\rm Tr}\left[S_{A}(x_{2},0)S(x_{2},y)X(y,0;t_{1},
\vec{q},\mu)\right],\label{eq:a} \\
B_{J_{\mu}}(t_{2},t_{1};\vec{p},\vpp,\Gamma) \equiv
\sum_{\vec{x}_{2},y}e^{-i\vec{p}\cdot\vec{x_{2}}}
{\rm Tr}\left[S_{B}(x_{2},0)S(x_{2},y)X(y,0;t_{1},
\vec{q},\mu)\right].\label{eq:b}
\end{eqnarray}
The \lq ${\rm Tr}
$\rq\, notation denotes a trace over both color and Dirac indices.

We have succeeded in rewriting the three-point functions
for the proton and the neutron in a very compact manner. However,
Eqs.~(\ref{eq:a}) and (\ref{eq:b}) make it very clear that these quantities
can not be calculated directly because of the presence of the $S(x_{2},y)$
propagators, since there are sums over both $\vec{x}_{2}$
and $\vec{y}$ present.
These equations also make it clear that there are two
remedies for this situation. One possibility is
to introduce a source to simulate the
current, contained in the $X(y,0;t_{1},\vec{q},\mu)$ factor and associated with
the $\vec{y}$ sum above. This technique, of course, is not specific to the
nucleon and works for any hadron field. However,
as one can see from Eq.~(\ref{eq:x}), this choice fixes the spatial momentum
transfer, $\vec{q}$, for a given set of quark propagators.
The other possibility is
to introduce a source to simulate the two quark lines,
contained in $S_{A}$ and $S_{B}$, which lead
to the final nucleon associated with the $\vec{x}_{2}$
sum. This technique {\it is} specific to the nucleon,
but leaves the spatial momentum transfer
free. Indeed, even the
choice of which operator to reconstruct is deferred until
the final analysis, so that the propagators calculated are also
useful in studying, for example, an axial current. In summary,
fixing the current allows one to do a survey of particles with
a given probe, whereas fixing the particle source allows one
to use a variety of probes on a single particle. We choose here
to fix the particle source since we are particularly interested
in trying to understand the $q^{2}$ dependence of the nucleon
form factors.

For this purpose we introduce $v_{A,B}^{T}(y;t_{2},\vec{p})$ such that
\begin{eqnarray}
A_{J_{\mu}}(t_{2},t_{1};\vec{p},\vpp,\Gamma) =
\sum_{y}{\rm Tr}\left[v_{A}^{T}(y;t_{2},\vec{p})
X(y,0;t_{1},\vec{q},\mu)\right],\label{eq:aju}\\
B_{J_{\mu}}(t_{2},t_{1};\vec{p},\vpp,\Gamma) =
\sum_{y}{\rm Tr}\left[v_{B}^{T}(y;t_{2},\vec{p})
X(y,0;t_{1},\vec{q},\mu)\right]\,,\label{eq:bju}
\end{eqnarray}
where the transpose is over both Dirac and color indices. Explicitly,
\begin{equation}
v_{A,B}^{T}(y;t_{2},\vec{p}\,)\equiv
\sum_{\vec{x_{2}}}e^{-i\vec{p}\cdot\vec{x_{2}}}
S_{A,B}(x_{2},0)S(x_{2},y)\,.\label{eq:vab}
\end{equation}
We now multiply on the right of Eq.~(\ref{eq:vab}) by
$M(y,x')$ and sum on $y$ to give
\begin{equation}
\sum_{y}v_{A,B}^{T}(y;t_{2},\vec{p}\,)M(y,x')=
e^{-i\vec{p}\cdot\vec{x}'}S_{A,B}(x',0)\,.\label{eq:mvab}
\end{equation}
Using the well-known relation
\begin{equation}
M^{\dagger}(x,y)=\gamma_{5}M(y,x)\gamma_{5}\,,
\end{equation}
where \lq $\dagger$\rq\, works in Dirac
and color space, one can show that Eq.~(\ref{eq:mvab}) leads to
\begin{equation}
\sum_{y}M(x',y)\gamma_{5}v_{A,B}^{*}(y;t_{2},\vec{p}\,)=
e^{i\vec{p}\cdot\vec{x}'}\gamma_{5}S_{A,B}^{\dagger}
(x',0)\,.\label{eq:mvabb}
\end{equation}
The right hand side of (\ref{eq:mvabb}) identifies the sources
\begin{equation}
b_{A,B}(x';t_{2},\vec{p}\,)\equiv
e^{i\vec{p}\cdot \vec{x}'}\gamma_{5}S_{A,B}^{\dagger}(x',0)\,,
\end{equation}
to be inserted in the matrix inverter.
Actually, for our purposes, it is more convenient to consider a
linear combinations of sources,
\begin{eqnarray}
b^{proton}=q_{u}b_{A} +q_{d}b_{B},\\
b^{neutron}=q_{d}b_{A} +q_{u}b_{B},
\end{eqnarray}
which give the desired proton and neutron three-point functions
directly. (See Eqs.~(\ref{eq:gpp}) and (\ref{eq:gnn}) above.)

The general method of using secondary or sequential sources
to perform spatial sums over intermediate lattice
operators was introduced in Refs.~12.

\subsection{Correlation Functions}
Although the behavior of Euclidean-time nucleon correlation functions
on the lattice is standard material\cite{mart1,terry1,terry2,lat1,lat2},
we summarize their properties for future reference below.

We deal with both protons and neutrons,
but for simplicity the following discussion will be for the proton only.
The two-point function defined in Eq.~(\ref{eq:two}) in the large
Euclidean time limit gives
\begin{equation}
G_{pp}(t;\vec{p},\Gamma) \stackrel{t\,\gg 1}{\longrightarrow}
N_{v}\sum_{s}e^{-Et}\Gamma_{\alpha'\alpha}\langle {\rm vac}|\chi_{\alpha}
(0)|\vec{p},s\rangle\langle\vec{p},s|\overline{\chi}_{\alpha'}
(0)|{\rm vac}\rangle,\label{eq:two2}
\end{equation}
where $N_{v}$\, is the number of spatial points in the lattice
volume. We write lattice and continuum completeness,
respectively, for fermionic states as
\begin{eqnarray}
&\displaystyle{\sum_{n,\vec{p},s}}|n,\vec{p},s\rangle\langle n,\vec{p},s|=I,\\
&\displaystyle{\sum_{n,s}}\int\frac{d^3p}{(2\pi)^{3}}\frac{m}{E}
|n,\vec{p},s)( n,\vec{p},s|=I.
\end{eqnarray}
The continuum limit
\begin{equation}
\frac{1}{V}\sum_{\vec{p}}\longrightarrow
\int\frac{d^3p}{(2\pi)^{3}},
\end{equation}
where $V=N_{v}a^{3}$, gives the correspondence between lattice
and continuum states:
\begin{equation}
|n,\vec{p},s\rangle\longrightarrow\left(\frac{m}{VE}\right)^{1/2}
|n,\vec{p},s).\label{eq:cor2}
\end{equation}
Use of
Eqs.~(\ref{eq:cor1}) and (\ref{eq:cor2}) then give
\begin{eqnarray}
&\langle {\rm vac}|\chi_{\alpha}(0)|\vec{p},s\rangle\longrightarrow
\displaystyle{\frac{a^3}{(2\kappa)^{3/2}}\left(\frac{m}{N_{v}E}\right)^{1/2}}
({\rm vac}|\chi_{\alpha}^{cont.}(0)|\vec{p},s),\\
&\langle \vec{p},s|\overline{\chi}_{\alpha'}(0)|{\rm vac}
\rangle\longrightarrow
\displaystyle{\frac{a^3}{(2\kappa)^{3/2}}\left(\frac{m}{N_{v}E}\right)^{1/2}}
(\vec{p},s|\overline{\chi}_{\alpha'}^{cont.}(0)|{\rm vac}).
\end{eqnarray}
$( {\rm vac}|\chi_{\alpha}^{cont.}(0)|\vec{p},s)$
and $( \vec{p},s|\overline{\chi}_{\alpha'}^{cont.}(0)|{\rm vac})$
are related as usual to the free spinors $u_{\alpha}(\vec{p},s)$
and $\overline{u}_{\alpha'}(\vec{p},s)$ by
\begin{eqnarray}
&( {\rm vac}|\chi_{\alpha}^{cont.}(0)|\vec{p},s)=
Z\,u_{\alpha}(\vec{p},s),\\
&( \vec{p},s|\overline{\chi}_{\alpha'}^{cont.}(0)|{\rm vac})=
Z^{*}\,\overline{u}_{\alpha'}(\vec{p},s),
\end{eqnarray}
where the unknown amplitude, $Z$, transforms
as a scalar and therefore can not depend upon
$\vec{p}$ or $s$. Now, using these results in Eq.~(\ref{eq:two2}),
one obtains
\begin{equation}
G_{pp}(t;\vec{p},\Gamma) \stackrel{t\,\gg 1}{\longrightarrow}
\frac{|Z|^{2}a^{6}}{(2\kappa)^{3}}\frac{m}{E}e^{-Et}
\,{\rm tr}\left[\Gamma\left(\frac{-i\not{\!p}+m}{2m}\right)\right]\,,
\end{equation}
where we have used the following relation
for free spinors:
\begin{eqnarray}
\displaystyle{\sum_{s}}u(\vec{p},s)\overline{u}(\vec{p},s)=
\frac{(-i\not{\!p}+m)}{2m}.
\end{eqnarray}
For example, with the choice
\begin{equation}
\Gamma_{4}\equiv \frac{1}{2}\left(\begin{array}{cc}
                 I & ~~~~0 \\
                 0 & ~~~~0
\end{array}\right),
\end{equation}
one obtains
\begin{equation}
G_{pp}(t;\vec{p},\Gamma_{4}) \stackrel{t\,\gg 1}{\longrightarrow}
\frac{E+m}{2E}\frac{|Z|^{2}a^{6}}{(2\kappa)^{3}}e^{-Et}.\label{eq:two3}
\end{equation}
For the proton three-point function, Eq.~(\ref{eq:pro}),
the large time limit is
\begin{eqnarray}
&G_{pJ_{\mu}p}(t_{2},t_{1};\vec{p},\vpp,\Gamma)
\stackrel{(t_{2}-t_{1}),t_{1}\,\gg 1}{\longrightarrow}
-iN_{v}^{2}\displaystyle{\sum_{s,s'}}e^{-E_{p}(t_{2}-t_{1})}
e^{-E_{p'}t_{1}} \nonumber \\
&\cdot\,\Gamma_{\alpha'\alpha}
\langle {\rm vac}|\chi_{\alpha}(0)|\vec{p},s\rangle
\langle\vec{p},s|J_{\mu}(0)|\vec{p}\,',s'\rangle
\langle\vec{p}\,',s'|\overline{\chi}_{\alpha'}(0)|{\rm vac}\rangle.
\end{eqnarray}
The lattice and continuum matrix elements of $J_{\mu}$ are related by
\begin{equation}
\langle\vec{p},s|J_{\mu}(0)|\vec{p}\,',s'\rangle\longrightarrow
\frac{1}{N_{v}}\left(\frac{m^{2}}{E_{p}E_{p'}}\right)^{1/2}
(\vec{p},s|J_{\mu}^{cont.}(0)|\vec{p}\,',s'),
\end{equation}
where ( $\sigma_{\mu\nu}=\frac{1}{2i}[\gamma_{\mu},\gamma_{\nu}]$
and $F_{1}$ and $F_{2}$ are real)
\begin{equation}
(\vec{p},s|J_{\mu}^{cont.}(0)|\vec{p}\,',s')=
i\,\overline{u}(\vec{p},s)\left(\gamma_{\mu}F_{1}-
\sigma_{\mu\nu}\frac{q_{\nu}}{2m}F_{2}\right)u(\vec{p}\,',s').
\end{equation}
Based on these forms, with the choices $\vec{p}=0$,
$\mu=4$ and $\Gamma=\Gamma_{4}$,
the proton three-point function yields
\begin{equation}
G_{pJ_{4}p}(t_{2},t_{1};0,-\vec{q},\Gamma_{4})
\stackrel{(t_{2}-t_{1}),t_{1}\,\gg 1}{\longrightarrow}
\frac{|Z|^{2}a^{6}}{(2\kappa)^{3}}e^{-m(t_{2}-t_{1})}
e^{-Et_{1}}\left(\frac{E+m}{2E}\right)G_{e}(q^{2}),\label{eq:3ele}
\end{equation}
where $G_{e}(q^{2})\equiv\left(F_{1}-\frac{q_{\mu}^{2}}{(2m)^{2}}F_{2}\right)$
is the electric form factor. Similarly, using
\begin{equation}
\Gamma_{k}\equiv \frac{1}{2}\left(\begin{array}{cc}
                 \sigma_{k}& ~~~~0 \\
                 0 & ~~~~0
\end{array}\right),
\end{equation}
one finds that with $\vec{p}=0$, $\mu=j$ and $\Gamma=\Gamma_{k}$
\begin{equation}
G_{pJ_{j}p}(t_{2},t_{1};0,-\vec{q},\Gamma_{k})
\stackrel{(t_{2}-t_{1}),t_{1}\,\gg 1}{\longrightarrow}
\frac{1}{2E}\frac{|Z|^{2}a^{6}}{(2\kappa)^{3}}
e^{-m(t_{2}-t_{1})}e^{-Et_{1}}
\epsilon_{jkl}q_{l}G_{m}(q^{2}),\label{eq:3mag}
\end{equation}
where $G_{m}\equiv\left(F_{1}+F_{2}\right)$ is the magnetic form factor.

Eqs.~(\ref{eq:two3}), (\ref{eq:3ele}), (\ref{eq:3mag})
provide the means of
calculating $G_{e}$ and $G_{m}$. We do not measure these correlation
functions directly, but instead analyze the ratio
\begin{equation}
G_{e}(q^{2})=\left(\frac{2E}{E+m}\right)^{1-\frac{t_{2}}{2t'}}
\left(\frac{G_{pJ_{4}p}(t_{2},t_{2}/2;0,-\vec{q},\Gamma_{4})}
{G_{pp}(t_{2};0,\Gamma_{4})}\right)
\left(\frac{G_{pp}(t';0,\Gamma_{4})}
{G_{pp}(t';\vec{q},\Gamma_{4})}\right)^{\frac{t_{2}}{2t'}}\,,\label{eq:ge}
\end{equation}
for $G_{e}$ and
\begin{equation}
G_{m}(q^{2})=
\frac{1}{2}\epsilon_{jkl}
\frac{2E}{q_{l}}\left(\frac{E+m}{2E}\right)^{\frac{t_{2}}{2t'}}
\left(\frac{G_{pJ_{j}p}(t_{2},t_{2}/2;0,-\vec{q},\Gamma_{k})}
{G_{pp}(t_{2};0,\Gamma_{4})}\right)
\left(\frac{G_{pp}(t';0,\Gamma_{4})}
{G_{pp}(t';\vec{q},\Gamma_{4})}\right)^{\frac{t_{2}}{2t'}}\,,\label{eq:gm}
\end{equation}
for $G_{m}$, which are seen to have smaller error bars. Notice that
Eq.~(\ref{eq:ge}) guarantees that we measure $G_{e}(0)=1$ except for
convergence errors and that all currents are located halfway between the
creation and annihilation time steps.
Also notice that the fractional powers are
present in order to provide a choice in the two point
functions used in order to allow single exponential behavior to develop
in $t'$. The identities
developed in Refs.~1 and 6 show that $G_{pp}(t;\vec{q},\Gamma_{4})$,
$G_{pJ_{4}p}(t_{2},t_{1};0,-\vec{q},\Gamma_{4})$ and
$G_{pJ_{j}p}(t_{2},t_{1};0,-\vec{q},\Gamma_{k})$ are real functions in the
ensemble average after evenness (Eqs.~(\ref{eq:two3}) and
and (\ref{eq:3ele})) or
oddness (Eq.~(\ref{eq:3mag})) in spatial momentum $\vec{q}$ has been enforced
by hand. Because the time interval between the initial
and final nucleon sources here is odd,
$t_{2}=15$, we define a {\it spatial} current operator at half-time steps
according to
\begin{equation}
J_{j}(\vec{x}_{1},t_{1}+\frac{1}{2})\equiv \frac{1}{2}\left(
J_{j}(\vec{x}_{1},t_{1})+J_{j}(\vec{x}_{1},t_{1}+1)\right)
\end{equation}
for use in (\ref{eq:3mag}). The time-nonlocal charge density operator,
$J_{4}$, is already naturally associated with half-time steps.

\section{RESULTS}
Our quenched configurations are $16^{3}\times 24$ and were calculated
using the Monte Carlo Cabibbo-Marinari pseudoheatbath\cite{cabbibo}.
The $SU(3)$ fundamental Wilson action was used with periodic boundary
conditions and $\beta=6.0$. The gauge field was thermalized for $5000$
sweeps from a cold start and $12$ configurations separated by at least
$1000$ sweeps were saved. Since our gauge configurations are taken from a
single Markov chain, if the resulting correlations between configurations are
sufficiently small then the configurations chosen for analysis
are effectively statistically independent.
The preferred way to test for autocorrelations is to bin consecutive
configurations; we have too few configurations ($12$)
for this to be a useful test.
A binning of a larger data set\cite{Gupta} on quenched
lattice at $\beta=6.2$
revealed no evidence of autocorrelations for two-point hadronic functions
when configurations were separated by 250-500 (multi-hit Metropolis) sweeps.
An examination of a number of observables as a function of
Monte Carlo sweep number in the quenched $\beta=5.9$ form-factor calculation
of Ref.~5 revealed no apparent auto-correlations
for a sample of $28$ configurations separated by $1000$ pseudo-heatbath sweeps.
The same sweep-number separation was chosen for calculations of semi-leptonic
form factors in Ref.~15; we have made this same conservative choice
in the present calculation to keep auto-correlations to acceptably low levels.

For the quarks we use periodic boundary conditions
in the spatial directions and \lq fixed\rq\, time boundary conditions,
which consist of setting the quark couplings across the time edge
to zero. The origin of all quark propagators was chosen to be at
lattice time site $5$;
the secondary zero-momentum nucleon source was
fixed at time site $20$. We expect that these positions are sufficiently
far from the lattice time boundaries to avoid nonvacuum contaminations.
All our results for proton form factors use the point interpolating fields,
Eqs.~(\ref{eq:int1}) and (\ref{eq:int2}), and similarly for the neutron.
We used the conditioned conjugate gradient technique for quark
propagator evaluation described in Ref.~16. For our convergence
criterion we demanded that the relative change
in the absolute sum of the squares of the
quark or secondary propagators be less than $5\times 10^{-5}$ over
$5$ iterations. As one check of the nucleon secondary source, we
verified current conservation for $t_{2} > t_{1} > 0$ to ${\it O}(10^{-4})$.
Since we wish to calculate the electric and magnetic form factors of
both the proton and the neutron, one non-source and four source propagator
inversions are necessary per configuration. The results below include
$\kappa=.154,.152,.148$, and $.140$. Our statistical
error bars come from first (form factors),
second (form factor fits) and third order (chiral extrapolations)
single-elimination jackknifes.

We show the pion, rho and nucleon masses measured on our $12$
configurations in Table I. (The interpolation fields used for the
pion and rho are the usual relativistic ones.)
Actually, we show the results of single
exponential mass fits using both smeared (over the entire lattice
volume using the lattice Coulomb gauge) and point quark propagators.
The fits here are for lattice time sites $16$ to $19$ for $\kappa=
.154$,$.152$ and $.148$ and time sites $18$ to $21$ for $\kappa=.140$,
which was seen to evolve more slowly in time. The smeared and point
masses are consistent with one another within the statistical error bars,
but a small systematic downward shift of the smeared
masses relative to the point masses seems to be present. The smeared
results are also consistent with the more accurate $\beta=6.0$ mass
results of Ref.~17, with which our largest two $\kappa$ values overlap.
When needed, the smeared mass
results from Table I (from our simulation) will be used for the kinematic
factors which appear in (\ref{eq:ge}) and (\ref{eq:gm});
the uncertainties associated the kinematics
are then included in the form factors as uncorrelated errors,
which affects mainly the magnetic error bars.
We will also use the accurately
determined $\beta=6.0$
value of $\kappa_{cr}=.15708(2)$ from Ref.~18 for our chiral extrapolations.
In the following, we will illustrate our data mainly with the
$\kappa=.154$ results, where, it must be kept in mind, the error bars are
the worst.

In order to test for continuum dispersion and to examine the time dependence
of our two point functions, we define the local mass, energy, and energy minus
mass from (\ref{eq:two3}) as
\begin{equation}
m(t+\frac{1}{2})={\rm ln}\left(\frac{G_{pp}(t;0,\Gamma_{4})}
{G_{pp}(t+1;0,\Gamma_{4})}\right)\, ,\label{eq:local1}
\end{equation}
\begin{equation}
E(t+\frac{1}{2})={\rm ln}\left(\frac{G_{pp}(t;\vec{p},\Gamma_{4})}
{G_{pp}(t+1;\vec{p},\Gamma_{4})}\right)\, ,\label{eq:local2}
\end{equation}
\begin{equation}
[E-m](t+\frac{1}{2})={\rm ln}\left(\frac{G_{pp}(t;\vec{p},\Gamma_{4})
G_{pp}(t+1;0,\Gamma_{4})}{G_{pp}(t;0,\Gamma_{4})
G_{pp}(t+1;\vec{p},\Gamma_{4})}\right)\, .\label{eq:eminusm}
\end{equation}
These quantities are given in Fig.~1 for $\kappa=.154$ as a function
of lattice time. The starting position of all quark propagators is time
step 5. The horizontal lines in this figure give the expected
result from the continuum dispersion relation using the central value of
the measured nucleon mass.
The most significant quantity relevant
to our form factor measurements is
Eq.~(\ref{eq:eminusm}) since this involves a
ratio of quantities that enters in Eqs.~(\ref{eq:ge}) and
(\ref{eq:gm}). From the figure, it appears that this measurement does
not become consistent with continuum dispersion until time slice
$14\frac{1}{2}$, which corresponds to $t'=9$ in Eqs.~(\ref{eq:ge}) and
(\ref{eq:gm}). (The propagator time origin is defined to be $t'=0$).
The other $\kappa$ values are similar, and $t'=9$
is used in all of our form factor results.

Figures 2 and 3 represent energy minus mass measurements as measured
from ratios of
the three-point functions, Eqs.~(\ref{eq:3ele}) and (\ref{eq:3mag}).
These are defined at integer time steps (the currents
at $t_{1}$ are
defined at half-integer sites) by
\begin{equation}
[E-m](t_{1}+\frac{1}{2})|_{j_{4}}={\rm ln}\left(\frac{
G_{pJ_{4}p}(t_{2},t_{1};0,-\vec{q},\Gamma_{4})}
{G_{pJ_{4}p}(t_{2},t_{1}+1;0,-\vec{q},\Gamma_{4})}\right)\, ,\label{eq:emm2}
\end{equation}
\begin{equation}
[E-m](t_{1}+\frac{1}{2})|_{j_{k}}={\rm ln}\left(
\frac{G_{pJ_{k}p}(t_{2},t_{1};0,-\vec{q},\Gamma_{j})}
{G_{pJ_{k}p}(t_{2},t_{1}+1;0,-\vec{q},\Gamma_{j})}
\right)\, .\label{eq:emm3}
\end{equation}
Figure 2 shows the local $[E-m]$ values from
the magnetic three-point function.
This function is seen to have a flat exponential behavior which begins
quite near the source origin.  Figure 3 shows the
$[E-m]$ values from the electric three-point function, which, unlike
the magnetic case, never appears to flatten.
On the other hand, the results
{\it are} consistent
with the expected
$[E-m]$ values near the mid-point of the lattice where the data for
Eqs.~(\ref{eq:ge}) and (\ref{eq:gm}) are actually taken.

The results of our form factor measurements are presented in Tables
\ref{table2} through \ref{table5}.
Figures 4 though 7
present the $\kappa=.154$ graphical results.
Notice that the $(qa)^2>0$
values given in the tables are affected
very little by the error bars in the nucleon mass
(the maximum error bar, at $\kappa=.154$, is approximately $3\%$
of the central value).
Our philosophy in comparing our results to experiment is
to look at the simplest phenomenological forms consistent
with the lattice data, and then to extrapolate the fit
parameters, rather than the individual form factor values,
to the chiral limit.
The solid and broken lines
in these graphs represent the best {\em simultaneous}
dipole and monopole fits,
respectively, of the combined
proton electric, magnetic and neutron magnetic
form factors. These are three parameter fits, giving the
fit dipole mass from
\begin{equation}
G_{e}^{D}(q^{2})=\frac{1}{\left(1+\frac{q^{2}}{m_{D}^{2}}
\right)^{2}}\, ,\label{eq:dip1}
\end{equation}
or the monopole mass from
\begin{equation}
G_{e}^{M}(q^{2})=\frac{1}{1+\frac{q^{2}}{m_{M}^{2}}}\, ,
\end{equation}
as well as the proton and neutron magnetic moments from the
forms,
\begin{equation}
G_{m}^{D}(q^{2})=\frac{G_{m}(0)}{\left(1+\frac{q^{2}}{m_{D}^{2}}
\right)^{2}}\, ,\label{eq:dip2}
\end{equation}
or
\begin{equation}
G_{m}^{M}(q^{2})=\frac{G_{m}(0)}{1+\frac{q^{2}}{m_{M}^{2}}}\, .
\end{equation}
The fit parameters found this way are listed in Table \ref{table6}
along with the chi-squared per degree of freedom found ($\chi^{2}_{d}$),
and the dipole/nucleon or monopole/nucleon
mass ratio.
In general, the quality of the fits
are reasonable except for the monopole form at $\kappa=.140$.
The simultaneous monopole fits are seen
to be slightly preferable to the dipole ones at the lowest three
$\kappa$ values. Notice also that the ratio $m_{M}/m_{N}$ is essentially
flat over these three $\kappa$ values.

In a separate fit of the proton electric form factors, we
list in Table \ref{table7} the dipole and monopole masses,
the corresponding charge radii (in lattice units),
mass ratios, and $\chi^{2}_{d}$ values found.
In comparing the $\chi^{2}_{d}$ values from the dipole fits in
Tables \ref{table6} and \ref{table7}, we notice that the
inclusion of the magnetic data decreases the quality of the
fits at $\kappa=.154$ and $.140$, whereas it increases the quality
at the two intermediate $\kappa$ values. For the monopole fits,
we see that the $\chi^{2}_{d}$ values are low in all cases,
except again for the $\kappa=.140$
simultaneous fit. It is the inclusion of
the magnetic data that is responsible for the large
$\chi^{2}_{d}$ value; the values in Table \ref{table5}
inform us that the $(qa)^{2}$ falloff of the
magnetic data is faster than for the proton electric data
at this $\kappa$ value.

Figures 7 through 10 show the measurements
of the neutron electric form factor at the four $\kappa$ values.
The phenomenological form,
\begin{equation}
G_{e}^{n}(q^{2})=-\frac{q^{2}}{4m_{N}^{2}}G_{m}^{n}(q^{2})\, ,
\label{eq:ph1}\end{equation}
is compared to the numerical results, using either
the dipole (solid line) or monopole (broken line) parameters from
Table \ref{table6} to characterize $G_{m}^{n}(q^{2})$. Although the
data are quite noisy, we obtain positive values of $G_{e}^{n}(q^{2})$
in all cases, in agreement with most experiments in this
energy regime. In addition, we
see that the above phenomenological
form fails at the two lowest $\kappa$ values. This does not
rule out such a form in the chiral limit, but does
make whatever physics lies behind it less compelling.

Figure 11 shows the $\kappa=.152$ proton electric form
factor. In our measurements we have the option of reconstructing the
spatial momentum transfers in a number of different directions for
a given $(qa)^{2}$ value. This figure shows the
effect on the error bars of averaging ($\Diamond$) and
and not averaging ($\Box$) over equivalent momenta. (The $(qa)^{2}$
values of the nonaveraged data have been increased slightly so that
the two data sets do not overlap.) At these values
of $\kappa$ the effect is to reduce the error bars
by a factor of two to three.

Also shown in Fig.~11 are the best
dipole (solid line) and monopole (broken line) fits to this
data from Table \ref{table7}.
These fits illustrate
that the source of the large $\chi^{2}_{d}$ value seen in the dipole
fit of Table \ref{table7} at $\kappa=.152$
(and similarly at $.148$) comes
about because of the failure of the highest $(qa)^{2}$ measurement to fall
off sufficiently fast. It is possible this is a systematic high $(qa)^{2}$
lattice artifact; on the other hand, the fact
that the $\kappa=.154$ and $.140$ results do not display
similar behavior undercuts this explanation.

Figure 12 represents a comparison of the results of
two methods of extracting the proton magnetic moment. In this
figure, the data points given by the square symbols are taken
from the dipole fits of Table \ref{table6}, while the diamond data points
are given by the zero momentum measurement ($t_{2}> t_{1}>0$):
\begin{equation}
\frac{\displaystyle{ \sum_{\vec{x}_{2},\vec{x}_{1}}}
	{\rm tr}\left[ \Gamma_{k} \langle{\rm vac}|
		\chi(x_{2})
	  (x_{1})_{i}J_{j}(x_{1})\overline{\chi}(0)|
{\rm vac}\rangle\right]}
{\displaystyle{\sum_{\vec{x_{2}}}}
{\rm tr}\left[\Gamma_{4}
\langle{\rm vac}|\chi(x_{2})
	\overline{\chi}(0)|{\rm vac}\rangle\right]}
= \epsilon_{ijk}\frac{G_{m}(0)}{2m}. \label{eq:pseudo}
\end{equation}
The continuum formula on which the above is based is derived
by taking the derivative
of the continuum analog of $G_{pJ_{j}p}(t_{2},t_{1};0,-\vec{q},\Gamma_{k})$
with respect to the $i$th component of $\vec{q}$, evaluated at $\vec{q}=0$,
and dividing by a zero momentum two point function. The resulting equation
is then transcribed into lattice language by
changing the continuum matrix elements into lattice ones and by making the
substitutions $\int d^{3}x\rightarrow a\sum_{\vec{x}}$ and
$J_{j}^{cont.}(\vec{x})\rightarrow a^{-3}J_{j}(\vec{x})$. Although the
two results agree at the smallest $\kappa$,
the zero momentum measurements give unrealistically small
magnetic moments at the larger $\kappa$ values.
The reason for this behavior is the same as for similar
behavior seen in lattice mesons using charge overlap techniques
\cite{me}. Because the lattice matrix elements do not contain arbitrarily
small momentum states, the continuum derivative at $\vec{q}=0$
can not be duplicated, and
the lattice version would only be
expected to hold for $D/2\gg R$, where $D$ is the length of the lattice
on one side and $R$ is a hadron correlation length, say, the charge radius.
That is, the hadron is expected to be well contained
in the given lattice volume. Apparently, this condition is only beginning
to be satisfied at the smallest measured $\kappa$ value, which is the
farthest from the chiral limit.

Figures 13 and 14
represent the chiral extrapolation of the dipole fit
proton and neutron magnetic moments. The values found from
these fits as well as from similar monopole fits are listed in Table
\ref{table8}. The quantity $m_{q}a$ is defined to be
\begin{equation}
m_{q}a\equiv \frac{1}{2}\left(\frac{1}{\kappa}-\frac{1}{\kappa_{cr}}
\right). \label{eq:mq}
\end{equation}
These extrapolations are simply linear fits,
which were adequate to describe the data, as seen from the
$\chi^{2}_{d}$ values. The magnitudes of the magnetic moments
are $10$ to $15\%$ low; the magnitudes found
in Ref.~5 are also low, but by $15$ to $30\%$.
There is a hint in these figures that the largest
$\kappa$ values prefer to lie above the linear fit, so more satisfactory
magnetic moments may result from a deeper exploration of the
region near $\kappa_{critical}$ where, however, error bars
more problematical.

Figure 15 presents the chiral extrapolation of the dipole
to nucleon mass ratio from Table \ref{table8}. This is assumed to
be linear as a function of $m_{q}a$; again, the $\chi_{d}^{2}$ values
in Tables \ref{table8} and \ref{table9} do not demand a more
sophisticated treatment.
We prefer to do the chiral
extrapolations on the mass {\em ratios}
from the above monopole and dipole fits because
the ratio of similar physical quantities is often less
subject to systematic errors and because
measuring the nucleon mass in relation to other
hadrons is best done in a separate high statistics
spectrum calculation, such as the $\beta=6.0$ calculations of
Refs.~17 and 18. For comparison, we have provided the
results of three parameter (Table \ref{table8})
and one parameter (Table \ref{table9})
fits of experimental nucleon data taken from Refs.~20 and 21.
Setting the scale from the chiral limit nucleon mass of
Ref.~18, our four-momentum transfer range roughly
corresponds to $.6~{\rm GeV}^{2}\leq q^{2}\leq 1.9~{\rm GeV}^{2}$;
the values overlapping with this range from
these references are the values used in the fits, which
are listed in Tables \ref{table10} and \ref{table11}.
The $\chi_{d}^{2}$ values for the
experimental results, which are more precise than the lattice
measurements, show that more parameters are really required
in this energy range to produce reasonable phenomenological fits.
We have $\chi_{d}^{2}= 1.99$ and $10.7$ for the dipole
and monopole fits, respectively,
of the proton electric data in Table \ref{table9}.
Similarly, we obtain $\chi_{d}^{2}= 3.39$ and $35.5$ for the
dipole and monopole fits, respectively, of Table \ref{table8}.
In comparing the two experimental fits,
we notice that the
dipole mass ratio is slightly larger when the magnetic
data is included (Table \ref{table8})
than when it is not (Table \ref{table9}).
Experiment shows that the proton and
neutron magnetic form factors fall off significantly more slowly
than the proton electric form factor in our energy regime\cite{gour}.
This explains the tendency of the experimental
dipole fit which includes the magnetic data
to produce a larger value of $m_{D}/m_{N}$ than
a similar fit of the proton electric data alone. It is
encouraging that the same tendency seems to be present in the
lattice dipole results in Tables \ref{table8}
and \ref{table9}, although the overall value for the ratio seems
to be about $7\%$ low in either case. However, as in the
magnetic moment case, the larger $\kappa$ values of this ratio prefer to
lie above the linear fit, and therefore
a more satisfactory value of this ratio could
result from a deeper exploration of the chiral limit.

\section{COMPARISON AND CONCLUSIONS}
We have investigated the functional forms
of the nucleon electromagnetic form factors as given by quenched
lattice QCD. Although our results
are not sufficiently accurate to distinguish
between monopole and dipole fits to the data, we have seen that
the error bars on these quantities are encouraging and that
the magnetic moments as well as the
dipole to nucleon mass ratio are reproduced to within about
$15\%$, similar to spectrum calculations. We have also seen
an indication that the chiral limit proton and neutron magnetic
form factors have a slower falloff in $q^{2}$
than the proton electric
form factor, which is similar to experiment in this energy regime.
In addition, the neutron electric form factors come out to be
positive, but their values are poorly
represented by a popular phenomenological form at intermediate to
small $\kappa$ values.
Finally, we
have investigated a zero-momentum technique for extracting
magnetic moments, but found that this method yields
unrealistically small values as $\kappa_{cr}$ is approached.

In agreement with the results of Ref.~5, which
used $12^{2}\times 24^{2}$ lattices at $\beta=5.9$, we have
found that the magnitudes of the proton and neutron
magnetic moments are small compared to experiment,
and, indeed, that the neutron value is more badly
represented than the proton.
The small improvement over the Ref.~5 values could
be due to the larger $\beta$ value used here. Alternatively, we have
seen a hint in Figs.~13 and 14 that perhaps better values
simply await a closer approach to $\kappa_{cr}$ rather
than requiring larger $\beta$ values. Another,
more interesting, possibility is the unknown contribution to
the magnetic moments from disconnected quark loops
due to current self-contractions\cite{lein}. The Ref.~5
calculation also gave positive values for the neutron electric
form factor as was found here.
There does seem to be a difference in the physical
size of the nucleons in these two studies, however. Corresponding to the
dipole and monopole chiral extrapolations in Table \ref{table9},
we find that that the dimensionless quantity $R_{p}m_{N}$
has a value $4.23(18)$ from the dipole fit and a value $4.80(35)$
from the monopole fit. If we simply divide these results by the
experimental (average) nucleon mass, equivalent to
the procedure adopted in Ref.~5, we then obtain a charge radius
of $.89(4){\rm fm}$ from the dipole fit and $1.01(7){\rm fm}$
from the monopole
fit as compared to the estimate $.65(8){\rm fm}$ from Ref.~5. The experimental
result is $.862(12){\rm fm}$\cite{simon}. The Ref.~5 result corresponds to a
dipole to nucleon mass ratio which is larger than the experimental
result by about $25\%$, whereas the result here is
about $7\%$ low compared to experiment. It is unlikely that the
physical box size plays an important role in this difference since,
assuming renormalization group scaling, the box dimensions
of these two calculations are very close (comparing
with the shortest box dimension of Ref.~5).
The systematics associated with the different techniques of
extracting the form factors
could be responsible for these rather different results.

Recent studies of scaling show that both the dimensionless ratios
of the string tension\cite{sv91} and
the scalar glueball mass to the chiral condensate\cite{sape89}
have scaling violations of $\sim 20 \%$ from $\beta = 5.7$ to
$\beta = 5.9$ and $\sim 10\%$ from $\beta = 5.9$ to $\beta = 6.0$.
Although glueball mass studies on large lattices seem to show
scaling from $ \beta = 5.9$ to $\beta = 6.2$ \cite{mt89},
hadron masses and $f_{\pi}$ ratios still show a deviation of the
order of $\sim 10$ to $20\%$ \cite{ggks91}. Therefore, masses and magnetic
moments measured at $\beta = 6.0$ are subject to a scale-breaking
systematic error which could be as large as $\sim 20 \%$, although
our use of mass ratios
in the monopole and dipole fits would be expected to significantly
reduce the systematic error in the
extrapolation of the functional forms to the chiral limit.

The overall message of form factor measurements to this point
seems to be that the quenched approximation
adequately represents the bulk of the physics of these quantities;
however, we are still far from being
able to {\em test} QCD in a precise experimental way on the lattice.
At the same time, we should keep in
mind that another major theme of such calculations
is the increased physics understanding that will
be afforded through
increasingly sophisticated parametrizations of lattice
laboratory data.
This has the potential of teaching us about
the dynamics associated with quark masses,
current self-contraction graphs,
and the quenched approximation.
It is clear that significantly larger computer resources will be
necessary to make substantial progress in our understanding
of these issues; we therefore look forward to the benefits
of improvements in computer technology, such as proposed in Ref.~29.

\section{ACKNOWLEDGEMENTS}
This research was supported in part by a DOE Grand Challenge Award,
DOE Grant No.~DE-FG05-84ER40154 as well as NSF Grant No.~STI-9108764. The
calculations were done on the CRAY~2 computers at the National Energy Research
Computer Center. W.W. would like to thank the
Baylor Summer Sabbatical Program for past support.
We are also very grateful to Dwight E. Neuenschwander and Chi-Min Wu for
assistance with the work described here and
to Weiqiang Liu for providing a copy of his Coulomb gauge-fixing program.

\newpage
\begin{table}
\caption{Smeared and point propagator mass fits.}\label{table1}
\begin{tabular}{ccccc}
\multicolumn{1}{c}{Particle} &\multicolumn{1}{c}{$\kappa=.154$}
&\multicolumn{1}{c}{$.152$} &\multicolumn{1}{c}{$.148$}
&\multicolumn{1}{c}{$.140$} \\
\tableline
\multicolumn{5}{c}{Smeared case} \\
pion   &  $.369(9)$ & $.488(7)$  &$.677(5)$ & $1.015(5)$\\
rho     & $.46(2)$  &  $.54(1)$  &$.711(9)$ & $1.032(5)$\\
nucleon & $.74(4)$  & $.85(2)$ &  $1.12(1)$ & $1.62(1)$\\
\multicolumn{5}{c}{Point case} \\
pion   &  $.38(2)$ & $.49(1)$  &$.684(9)$ & $1.027(6)$\\
rho     & $.46(2)$  &  $.55(1)$  &$.721(8)$ & $1.040(7)$\\
nucleon & $.73(5)$  & $.87(3)$ &  $1.15(2)$ & $1.64(1)$\\
\multicolumn{5}{c}{Ref.~17} \\
pion   &  $.361(1)$ & $.474(1)$  &$ - $ & $ - $\\
rho     & $.463(3)$  &  $.545(2)$  &$ - $ & $ - $\\
nucleon & $.721(7)$  & $.861(5)$ &  $ - $ & $ - $\\
\end{tabular}
\end{table}
\begin{table}
\caption{Proton and neutron form factors at $\kappa=.154$
as a function of lattice four momentum-transfer squared.}\label{table2}
\begin{tabular}{ccccc}
\multicolumn{1}{c}{Form Factor} &\multicolumn{1}{c}{$(qa)^{2}=.145$}
&\multicolumn{1}{c}{$.274$} &\multicolumn{1}{c}{$.392$}
&\multicolumn{1}{c}{$.502$} \\
\tableline
$G_{e}^{p}$   &  $.481(34)$ & $.335(34)$  &$.223(34)$ & $.186(59)$\\
$G_{m}^{p}$     & $1.17(11)$  &  $.895(90)$  &$.69(14)$ & $.86(24)$\\
$G_{e}^{n}$ & $.033(17)$  & $.026(17)$ &  $.046(38)$ & $.068(67)$\\
$G_{m}^{n}$   &  $-.748(69)$ & $-.578(89)$  &$-.526(88)$ & $-.54(20)$\\
\end{tabular}
\end{table}
\newpage
\begin{table}
\caption{Proton and neutron form factors at $\kappa=.152$
as a function of lattice four momentum-transfer squared.}\label{table3}
\begin{tabular}{ccccc}
\multicolumn{1}{c}{Form Factor} &\multicolumn{1}{c}{$(qa)^{2}=.147$}
&\multicolumn{1}{c}{$.281$} &\multicolumn{1}{c}{$.406$}
&\multicolumn{1}{c}{$.522$} \\
\tableline
$G_{e}^{p}$   &  $.551(18)$ & $.391(24)$  &$.293(29)$ & $.288(32)$\\
$G_{m}^{p}$     & $1.22(7)$  &  $.906(59)$  &$.696(95)$ & $.65(13)$\\
$G_{e}^{n}$ & $.0230(77)$  & $.019(11)$ &  $.026(20)$ & $.038(29)$\\
$G_{m}^{n}$   &  $-.781(59)$ & $-.586(47)$  &$-.474(49)$ & $-.406(90)$\\
\end{tabular}
\end{table}
\begin{table}
\caption{Proton and neutron form factors at $\kappa=.148$
as a function of lattice four momentum-transfer squared.}\label{table4}
\begin{tabular}{ccccc}
\multicolumn{1}{c}{Form Factor} &\multicolumn{1}{c}{$(qa)^{2}=.150$}
&\multicolumn{1}{c}{$.291$} &\multicolumn{1}{c}{$.426$}
&\multicolumn{1}{c}{$.555$} \\
\tableline
$G_{e}^{p}$   &  $.665(10)$ & $.496(17)$  &$.390(25)$ & $.367(27)$\\
$G_{m}^{p}$     & $1.42(4)$  &  $1.03(5)$  &$.795(66)$ & $.643(77)$\\
$G_{e}^{n}$ & $.0109(31)$  & $.0116(57)$ &  $.0129(81)$ & $.016(10)$\\
$G_{m}^{n}$   &  $-.907(36)$ & $-.666(32)$  &$-.520(39)$ & $-.403(49)$\\
\end{tabular}
\end{table}
\begin{table}
\caption{Proton and neutron form factors at $\kappa=.140$
as a function of lattice four-momentum transfer squared.}\label{table5}
\begin{tabular}{ccccc}
\multicolumn{1}{c}{Form Factor} &\multicolumn{1}{c}{$(qa)^{2}=.152$}
&\multicolumn{1}{c}{$.300$} &\multicolumn{1}{c}{$.444$}
&\multicolumn{1}{c}{$.584$} \\
\tableline
$G_{e}^{p}$   &  $.791(8)$ & $.646(14)$  &$.539(18)$ & $.484(21)$\\
$G_{m}^{p}$     & $1.44(3)$  &  $1.12(3)$  &$.895(41)$ & $.740(37)$\\
$G_{e}^{n}$ & $.0030(10)$  & $.0043(16)$ &  $.0048(22)$ & $.041(28)$\\
$G_{m}^{n}$   &  $-.927(22)$ & $-.726(25)$  &$-.582(30)$ & $-.476(22)$\\
\end{tabular}
\end{table}
\newpage
\begin{table}
\caption{Combined dipole and monopole
fits of the proton electric and magnetic,
neutron magnetic form factors.}\label{table6}
\begin{tabular}{c|ccccc}
   & & $\kappa=.154$ & $.152$ & $.148$ & $.140$ \\ \hline
Dipole & $m_{D}a$ & $.609(31)$ & $.688(21)$ & $.832(18)$ & $1.09(3)$ \\
             & $m_{D}/m_{N}$ & $.823(61)$ & $.809(31)$ & $.743(17)$
  & $.675(19)$ \\
  & $G_{m}^{p}(0)$ & $2.51(18)$ & $2.22(10)$ & $2.09(6)$ & $1.77(5)$ \\
  & $G_{m}^{n}(0)$ & $-1.59(13)$ & $-1.45(10)$ & $-1.34(5)$ & $-1.14(4)$ \\
  & $\chi^{2}_{d}$     & $1.36$ & $1.45$ & $.66$ & $1.63$ \\ \hline
Monopole & $m_{M}a$ & $.364(31)$ & $.427(19)$ & $.539(16)$ & $.725(24)$ \\
  & $m_{M}/m_{N}$ & $.492(49)$ & $.503(26)$ & $.481(15)$
  & $.447(15)$ \\
  & $G_{m}^{p}(0)$ & $2.63(22)$ & $2.26(10)$ & $2.09(6)$ & $1.77(6)$ \\
  & $G_{m}^{n}(0)$ & $-1.68(18)$ & $-1.46(11)$ & $-1.33(5)$ & $-1.13(5)$ \\
  & $\chi^{2}_{d}$     & $.70$ & $.25$ & $.64$ & $3.57$ \\
\end{tabular}
\end{table}
\begin{table}
\caption{Dipole and monopole
fits of the proton electric form factor.}\label{table7}
\begin{tabular}{c|ccccc}
   & & $\kappa=.154$ & $.152$ & $.148$ & $.140$ \\ \hline
 Dipole  & $m_{D}a$ & $.593(32)$ & $.679(20)$ & $.832(16)$ & $1.12(3)$ \\
   & $R_{p}/a$    & $5.84(32)$ & $5.10(15)$ & $4.16(8)$ & $3.09(8)$  \\
   & $m_{D}/m_{N}$ & $.801(61)$ & $.799(30)$ & $.743(16)$
& $.688(19)$ \\
  &  $\chi^{2}_{d}$ & $.25$ & $2.24$ & $1.97$ & $.41$ \\ \hline
 Monopole & $m_{M}a$ & $.359(31)$ & $.426(18)$ & $.542(16)$ & $.744(23)$ \\
  & $R_{p}/a$    & $6.82(59)$ & $5.75(24)$ & $4.52(13)$ & $3.29(10)$  \\
  & $m_{M}/m_{N}$ & $.485(49)$ & $.501(25)$ & $.484(15)$
  & $.459(15)$ \\
  &  $\chi^{2}_{d}$ & $.30$ & $.41$ & $.42$ & $.49$ \\
\end{tabular}
\end{table}
\newpage
\begin{table}
\caption{Chiral extrapolation of the proton, neutron magnetic moments
and the mass ratios from Table~\ref{table6}. The experimental
data is from Tables~\ref{table10} and \ref{table11}.}\label{table8}
\begin{tabular}{c|ccccc}
  & & value & $\chi^{2}_{d}$ & Exp'l fit & \\ \hline
  & $m_{D}/m_{N}$ & $.836(31)$ & $.39$ & $.900(5)$ & \\
 Dipole & $G_{m}^{p}(0)$ & $2.44(11)$ & $.64$ & $2.87(4)$ & \\
  & $G_{m}^{n}(0)$ & $-1.59(11)$ & $.23$ & $-2.02(5)$ & \\ \hline
 & $m_{M}/m_{N}$ & $.516(31)$ & $.06$ & $.448(4)$ & \\
 Monopole & $G_{m}^{p}(0)$ & $2.48(10)$ & $.81$ & $3.22(6)$ & \\
 & $G_{m}^{n}(0)$ & $-1.59(12)$ & $.51$ & $-2.33(6)$ & \\
\end{tabular}
\end{table}
\begin{table}
\caption{Chiral extrapolation of the dipole and monopole proton
mass ratios from Table~\ref{table7}.}\label{table9}
\begin{tabular}{cccccc}
  & & value & $\chi^{2}_{d}$ & Exp'l fit & \\ \hline
  & $m_{D}/m_{N}$ & $.818(35)$ & $.31$ & $.883(6)$ & \\
 & $m_{M}/m_{N}$ & $.510(37)$ & $.08$ & $.472(4)$ & \\
\end{tabular}
\end{table}
\begin{table}
\caption{Experimental data for the proton from Ref.~20.}
\label{table10}
\begin{tabular}{ccccc}
 & $q^{2}$ [$(GeV/c)^{2}$]  & $G_{e}(q^{2})$ & $G_{m}(q^{2})$ & \\ \hline
 & $.65$ & $.265(12)$ & $.767(8)$ & \\
 & $.72$ & $.270(17)$ & $.690(10)$ & \\
 & $.78$ & $.217(7)$ & $.647(6)$ & \\
 & $.94$ & $.196(8)$ & $.523(6)$ & \\
 & $1.1$ & $.141(5)$ & $.452(4)$ & \\
 & $1.35$ & $.114(5)$ & $.352(3)$ & \\
 & $1.75$ & $.0713(64)$ & $.248(3)$ & \\
\end{tabular}
\end{table}
\begin{table}
\caption{Experimental data for the neutron from Ref.~21.}
\label{table11}
\begin{tabular}{cccc}
 & $q^{2}$ [$(GeV/c)^{2}$]  & $G_{m}(q^{2})$ & \\ \hline
 & $.60$ & $-.629(20)$ & \\
 & $.78$ & $-.434(23)$ & \\
 & $1.0$ & $-.345(27)$ & \\
 & $1.0$ & $-.322(18)$ & \\
 & $1.17$ & $-.284(26)$ & \\
 & $1.53$ & $-.203(11)$ & \\
 & $1.80$ & $-.185(17)$ & \\
\end{tabular}
\end{table}
\pagebreak
\begin{center} {\bf Figure Captions} \end{center}

\begin{enumerate}

\item   Local energy, mass or energy minus mass for the nucleon two point
function, Eqs.~(\ref{eq:local1}) through (\ref{eq:eminusm}) for $\kappa=
.154$ as a function of lattice time slice.
In this and the next two figures,
the horizontal lines give the expected continuum results based upon
the measured mass. All propagators begin at time slice $5$.

\item   Local energy minus mass for the proton magnetic three point function,
Eq.~(\ref{eq:emm3}) at $\kappa=.154$ as a function of lattice time slice.

\item   Local energy minus mass for the proton electric three point function,
Eq.~(\ref{eq:emm2}) at $\kappa=.154$ as a function of lattice time slice.

\item   Electric form factor of the proton at $\kappa=.154$ as a function
of lattice Minkowski four momentum transfer squared. The solid
and broken lines represent
the best {\em simultaneous} dipole and monopole fits, respectively, of
the proton electric, magnetic and neutron magnetic form factors.
The parameters of the fits are listed in Table \ref{table6}.

\item   Magnetic form factor of the proton at $\kappa=.154$ as a function
of lattice Minkowski four momentum transfer squared. The solid
and broken lines are simultaneous dipole
and monopole fits, respectively,
as explained in the Fig.~4 caption.

\item   Magnetic form factor of the neutron at $\kappa=.154$ as a function
of lattice Minkowski four momentum transfer squared. The solid
and broken lines are simultaneous dipole
and monopole fits, respectively,
as explained in the Fig.~4 caption.

\item   Electric form factor of the neutron at $\kappa=.154$ as a function
of lattice Minkowski four momentum transfer squared. The lines shown
represent the phenomenological form, Eq.~(\ref{eq:ph1}),
using either dipole (solid line) or monopole (broken line) fits of
the neutron magnetic form factor from Table~\ref{table6}.

\item   Same as Fig.~7, except at $\kappa=.152$.

\item   Same as Fig.~7, except at $\kappa=.148$.

\item   Same as Fig.~7, except at $\kappa=.140$.

\item   A comparison at $\kappa=.152$ of the proton electric
form factors calculated by averaging all equivalent momentum
transfers ($\Diamond$) and by not averaging ($\Box$).
The solid
and broken lines represent
the best dipole and monopole fits of the proton electric factor,
respectively, from Table \ref{table7}.

\item   The Magnetic moment of the proton at four values of
$m_{q}$, which is defined by Eq.~(\ref{eq:mq}).
The values from the simultaneous
fits given in Table \ref{table6} ($\Box$) are contrasted with
the extracted values from Eq.~(\ref{eq:pseudo}) ($\Diamond$).

\item   Linear chiral extrapolation
in $m_{q}$ of the proton magnetic
moment from the simultaneous fits of Table \ref{table6}.

\item   Linear chiral extrapolation
in $m_{q}$ of the neutron magnetic
moment from the simultaneous fits of Table \ref{table6}.

\item   Linear chiral extrapolation
in $m_{q}$ of the dipole to nucleon mass ratio
from the simultaneous fits of Table \ref{table6}.

\end{enumerate}

\begin{references}
\bibitem[11]{explain} We require the correspondence
$$e^{iS_F^{M}[\overline\psi,\psi]}\rightarrow
e^{-S_F^{E}[\overline\zeta,\zeta]}$$
under $t\rightarrow -it$, which identifies $S_F^{E}[\overline\zeta,\zeta]$ as
$$S_F^{E}[\overline\zeta,\zeta]=
\sum_{I,J}\overline\zeta_{I}M_{IJ}[U^{\dagger}]\zeta_{J}\,,$$ where
$U_{\mu}^{\dagger}$ is now the usual Euclidean link variable.
\end{references}
\end{document}